\documentclass[showpacs,twocolumn,pra]{revtex4}
\usepackage{txfonts}
\usepackage{amssymb}

\usepackage{graphicx}
\usepackage{dcolumn}
\usepackage{bm}

\begin{document}

\title{Approach to accurately measuring the speed of optical precursors}
\author{Chuan-Feng Li,$^{1}\footnote{
email: cfli@ustc.edu.cn}$ Zong-Quan Zhou,$^{1}$ Heejeong Jeong,$^{2}$ and Guang-Can Guo$^{1}$} \affiliation{ $^1$Key Laboratory of Quantum
Information, University of Science and Technology of China, CAS, Hefei, 230026, People's Republic of China
\\$^2$Thayer School of Engineering, Dartmouth College, Hanover, New
Hampshire, 03755, USA}
\date{\today }

\begin{abstract}
Precursors can serve as a bound on the speed of information with dispersive medium. We propose a
method to identify the speed of optical wavefronts using polarization-based interference in a
solid-state device, which can bound the accuracy of the speed of wavefronts to less than $10^{-4}$
with conventional experimental conditions. Our proposal may have important implications for optical
communications and fast information processing.
\end{abstract}

\pacs{42.50.Gy 42.25.Bs 03.30.+p}
\maketitle

The advent of fast information processing in quantum network devices often raises the need for accurate
control of light pulse propagation. The speed of the information encoded on the optical pulse
should obey the causality \cite{Einstein1905}. This seemingly obvious fact has been experimentally
proved only recently \cite{Gauthier03}. Unlike the well-defined velocity of a point particle, the
velocity of a light pulse traveling through an optical material is not precisely defined. The
motion of a light pulse can be approximated by the group velocity ($\upsilon_{g}$), which is given
by $\upsilon_{g}=c/(n+\omega dn/d\omega), $ where $c$ is the speed of light in a vacuum, $n$ is the
refractive index of the material and $\omega$ is the frequency of the light \cite{Boyd09}. It can
be clearly seen that $\upsilon_{g}$ can be greater than $c$ and even be negative when $\omega
dn/d\omega$ is negative (anomalous dispersion). Experimental studies of fast light propagation
include Refs. \cite{Wang00,Boyd03,Gauthier03,Dolling06,Boyd06}. To resolve the apparent contradictions
between fast light propagation and the theory of relativity, optical precursors were introduced by
Sommerfeld and Brillouin in 1914 \cite{Sommerfeld1914,Brillouin1914,Brillouin1960}. The theory
states that the front edges of an ideal step-modulated pulse propagate at speed $c$ because of
the finite response time of any physical material and no components can overtake this wave front.
The forerunners, now known as Sommerfeld-Brillouin precursors, are followed by the main pulse
traveling at its group velocity. This conclusion, while conceptually clear, lacks of experimental
evidence directly relating to the speed of optical precursors.

Optical precursors are recently of great interest because they have applications in biomedical
imaging \cite{Albanese89}, underwater communications \cite{Choi04}, and the generation of high peak
power optical pulses \cite{Jeong10,Du10}. A number of theoretical and experimental studies have
been carried out \cite{Jeong06,Aaviksoo91,Theory09,Falcon03,Du08,Du09,Du11}. Specifically, direct
observation of precursors which are separated from the delayed main fields has been achieved with
electromagnetically induced transparency (EIT) in cold atoms \cite{Du09,Du11}. For all these works,
the authors claim that the rising edge of the input pulse propagates at the speed of $c$ in the
form of Sommerfeld precursors. However, considering the rise time of the detecting systems, a time
delay of shorter than 1 ps would not be captured if the precursors travel at a speed which differs
little from $c$.

In this paper, we propose an experimental method to precisely determine the speed of
optical wavefronts propagating through a dispersive medium, i.e., optically pumped
Nd$^{3+}$:YVO$_{4}$ crystal by interference with a reference pulse traveling through a pure
YVO$_{4}$ crystal. If there is any speed difference between wavefronts in the Nd$^{3+}$:YVO$_{4}$
crystal and the YVO$_{4}$ crystal, the interference pattern will be significantly altered and reveal
the influence of Nd$^{3+}$ ions on the speed of the wavefronts.

Nd$^{3+}$:YVO$_{4}$ crystals (doping level, 10 ppm.) have been systematically investigated as a
candidate solid-state quantum memory \cite{Gisin08n,Gisin08,Gisin10}. The $^{4}I_{9/2}\rightarrow{
}^4F_{3/2}$ transition of Nd$^{3+}$ around 879.705 nm has a narrow homogenous linewidth
($\Gamma_{h}\sim$63 kHz) and wideband inhomogeneous broadening ($\Gamma_{inh}\sim$2.1 GHz) at low
temperature. Nd$^{3+}$ is a Kramers ion and thus has strong first-order Zeeman effect, which
yields a large ground-state splitting under the application of a magnetic field (10-100
GHz/T). From the inset of Fig. 1, it can be seen that a moderate magnetic field splits the ground
states into two Zeeman spin levels ($|1\rangle,|2\rangle$) which connect to an excited state
($|3\rangle$). Thus, this provides a $\Lambda$-like system \cite{Gisin08,Gisin10}.

Note that any real pulse has a step-rising front which is composed of the infinite spectral
components, even the pure YVO$_{4}$ crystal cannot response to it, the ideal step-rising wave front
propagates at the speed of $c$ independent of any medium. While for the light pulse produced by
electro-optic modulator (EOM) with a finite rise time, the YVO$_{4}$ crystal behaves as a wideband
response medium, the wave fronts which travels at the speed of $c$ are generally too weak for detection in the YVO$_{4}$
crystal. In both the Nd$^{3+}$:YVO$_{4}$ crystal and the pure YVO$_{4}$ crystal the experimentally
detectable wavefront should propagate at the same speed $\upsilon_{0}=c/n_{0}$, where $n_0$ is the
refractive index for far-detuned frequency components of the input pulse. The
refractive index ideally reaches unity and $\upsilon_{0}$ becomes $c$ for infinite frequency
components. In the Nd$^{3+}$:YVO$_{4}$ crystal, the optically pumped Nd$^{3+}$ ions have a strong
dispersion of the narrow-band frequency components that is near the resonance line of the
Nd$^{3+}$. Those components make up the greatly delayed or accelerated main pulse.

\begin{figure}[tbph]
\begin{center}
\includegraphics[width= 3.1in]{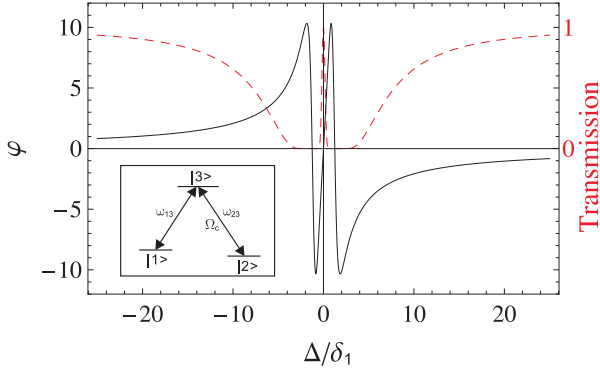}
\end{center}
\caption{ (Color online. The phase (black solid line) and intensity (red dashed line) of light
after passing through a Nd$^{3+}$:YVO$_{4}$ crystal versus the probe detuning. The inset is an energy
level diagram of the $^{4}I_{9/2}\rightarrow{ }^4F_{3/2}$ transition of Nd$^{3+}$ under a magnetic
field.} \end{figure}

The frequency of the probe light is tuned to the Nd$^{3+}$ absorption line ($\omega_{31}$). It is a
square pulse modulated using an EOM with a temporal width of $t_{0}=4\times10^{-6}$ s and a rise
(fall) time of $t_{r}=0.1\times10^{-9}$ s. The input light intensity is assumed to be unity in the
following calculations. $E(0,\omega)$ is the Fourier transform of the input pulse. The
Nd$^{3+}$:YVO$_{4}$ crystal undergoing EIT is characterized by its linear susceptibility
\cite{EIT05}
\begin{eqnarray}
\chi_{1}(\omega)=\frac{\alpha_{0}}{k_{0}}\frac{4(\Delta+i\gamma_{12})\delta_{1}}{|\Omega_{c}|^{2}-4(\Delta+i\gamma_{12})(\Delta+i\delta_{1})},
\end{eqnarray} where $\alpha_{0}=4130$ m$^{-1}$ is the on-resonance absorption coefficient;
 $k_{0}=n_{0}\omega_{31}/c= 1.55\times10^{7}$ m$^{-1}$; $n_{0}$ is the refractive index of the
YVO$_{4}$ crystal near 879.705 nm for light polarized parallel to the crystal's symmetric $c$ axis
\cite{index}; $\Delta=\omega-\omega_{31}$ is the probe detuning;
 and $\Omega_{c}$ is the coupling laser Rabi frequency. 2$\delta_{1}$ is the original absorption bandwidth, which is controlled by the pumping procedure.
Here, we choose $\delta_{1}=2\pi\times$1 MHz which is larger than $\Gamma_{h}$. The dephasing rate
between the two ground states is $\gamma_{12}=2\pi\times$0.2 kHz. These parameters were chosen
based on experimental observations \cite{Gisin08,Gisin10}. The transmitted field is given by
$E(z_{1},\omega)=E(0,\omega)Exp[ik_{1}(\omega)z_{1}]$ with
$k_{1}(\omega)=k_{0}\sqrt{1+\chi_{1}(\omega)}\simeq k_{0}(1+\chi_{1}(\omega)/2)$. Via a fast
Fourier transformation (FFT), the transmitted field in time domain can be obtained. There have been
some analytical solutions to this problem \cite{Theory09}, which fit well with the results obtained
using the FFT. However, to preserve the phase of the field for the coming interference experiments,
we use FFT to evaluate the integrals. Fig. 1 shows the phase of the photons experiencing
$\varphi=Re[k_{1}(\omega)z_{1}]-k_{0}z_{1}$ with $z_{1}=0.01$ m and the transmitted-field intensity of
$Exp[-2Im[k_{1}(\omega)z_{1}]]$. It can be seen that far-detuned light experiences a smaller phase
shift than near-resonance light. After passing through the EIT medium, the rising and falling edges at $t=0$ and $t=t_{0}$ should not experience group delay caused by Nd$^{3+}$ ions.
However, the main fields are approximately delayed by
$t_{g}=2\alpha_{0}z_{1}\delta_{1}/|\Omega_{c}|^{2}$ \cite{Theory09, Du09}.

\begin{figure}[tbph]
\begin{center}
\includegraphics[width= 3.1in]{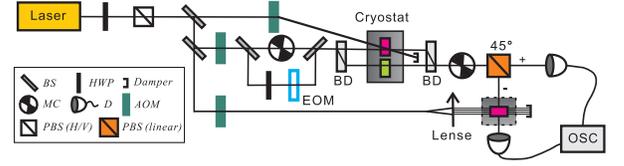}
\end{center}
\caption{ (Color online) Experimental setup for measuring the speed of optical wavefronts. The
coupling light produced by the top AOM should overlap with the probe light in the first
Nd$^{3+}$:YVO$_{4}$ crystal ( top, red filled) and be damped then. The pump light for the the first
Nd$^{3+}$:YVO$_{4}$ crystal is produced by the middle of the AOM and it is controlled by a MC. The
input pulse is produced by the EOM. The lower AOM produces the pump light for the filter. The pure
YVO$_{4}$ crystal (bottom, green filled) is used as a reference path. After passing through the BD,
the light undergoes polarization-dependent detection which is recorded on an oscilloscope (OSC).}
\end{figure}

The experimental scheme is shown in Fig. 2. The laser should be wavelength tunable and have a
linewidth well below 1 MHz. Using a polarization beam splitter (PBS) and a half-wave plate (HWP),
the light's polarization can be tuned to maximal absorption for a Nd$^{3+}$:YVO$_{4}$ crystal. The
top acousto-optical modulators (AOM) shift the laser frequency to be on-resonance with
$\omega_{23}$ to serve as the coupling light. The middle of the AOM sweep of the laser frequency
around $\omega_{13}$ serves as the pump light. The mechanical chopper (MC) before the BD turns the
strong pump light off during the detection cycle. The input pulse is generated using a 15-GHz EOM
and the light's polarization is rotated to $|+\rangle=\frac{1}{\sqrt{2}}(|H\rangle+|V\rangle)$ by the
HWP, where $H$ and $V$ denote horizontal and vertical polarization state, respectively. To show that the wavefronts
actually do travel at the speed of $\upsilon_{0}$ independently of the Nd$^{3+}$ ions, we use a
reference pulse traveling through a pure YVO$_{4}$ crystal. The beam displacer (BD) separates $H$ and
$V$ polarized light. The $H$ polarized probe light is sent through the Nd$^{3+}$:YVO$_{4}$ crystal
(top, red filled) such that it overlaps with coupling light at a small angle. The $V$ polarized probe
light is sent through the pure YVO$_{4}$ crystal (bottom, green filled) of the same length, $z_{1}$.
The Nd$^{3+}$:YVO$_{4}$ crystal's $c$ axis is aligned in the horizontal direction and the YVO$_{4}$
crystal's $c$ axis is aligned in the vertical direction to avoid birefringence. All of the crystals
are placed in a cryostat with a temperature 3 K and a magnetic field 0.5 T. Then $H$
and $V$ polarized light are combined using the BD and then undergo polarization dependent detection.
The MC after the BD is used to block light to protect the filter and detectors during the pumping
procedure of the first Nd$^{3+}$:YVO$_{4}$ crystal. The transmitted light intensities for the
$|+\rangle$ and $|-\rangle=\frac{1}{\sqrt{2}}(|H\rangle-|V\rangle)$ polarizations are recorded
using an oscilloscope (OSC). The intensities are given by
\begin{eqnarray}
I_{+}(z_{1},t)=|IFT\{\frac{1}{\sqrt{2}}[E_{H}(z_{1},\omega)+E_{V}(z_{1},\omega)]\}|^{2}
\\I_{-}(z_{1},t)=|IFT\{\frac{1}{\sqrt{2}}[E_{H}(z_{1},\omega)-E_{V}(z_{1},\omega)]\}|^{2},
\end{eqnarray} where
\begin{eqnarray}
E_{H}(z_{1},\omega)=\sqrt{1/2}E(0,\omega)Exp[ik_{1}(\omega)z_{1}],
\\E_{V}(z_{1},\omega)=\sqrt{1/2}E(0,\omega)Exp[ik_{0}z_{1}]
\end{eqnarray} and IFT means Inverse Fourier Transformation.
Note that the phase velocity is the relevant quantity for interference, which is the same in a
pure crystal and a low-doping crystal. If the $H$ polarized field and $V$ polarized field arrive at the same time, the
interference shall give an ideal near-to-zero output for $I_{-}$ in the first nanoseconds.

To totally eliminate the main fields with the central frequency components, which experience
greater phase shifts, another Nd$^{3+}$:YVO$_{4}$ crystal with length $z_{2}$ is placed before
the detectors for $|-\rangle$ polarized light. The crystal's $c$ axis is aligned in the $-45^{\circ}$
direction for maximal absorption. It acts as a strong absorption filter which can be characterized
by
\begin{eqnarray}
\chi_{2}(\omega)=\frac{\alpha_{0}}{k_{0}}\frac{\delta_{2}}{-(\Delta+i\delta_{2})}.
\end{eqnarray}
This is the special case of Eq. (1) with $\Omega_{c}=0$. $\delta_{2}$ is controlled by the pump light
produced by the lower AOM in Fig. 2. In the following text, we refer to this Nd$^{3+}$:YVO$_{4}$
crystal as the ``filter" for simplicity. With the filter present, the recorded intensity for $I_{-}$ is
given by
\begin{eqnarray}
I_{-}(z_{2},t)=|IFT\{\frac{1}{\sqrt{2}}[E_{H}(z_{1},\omega)-E_{V}(z_{1},\omega)]e^{ik_{2}(\omega)z_{2}}\}\mid^{2},
\end{eqnarray} where $k_{2}(\omega)\simeq k_{0}(1+\chi_{2}(\omega)/2)$.

\begin{figure}[tbph]
\begin{center}
\includegraphics[width= 2.9in]{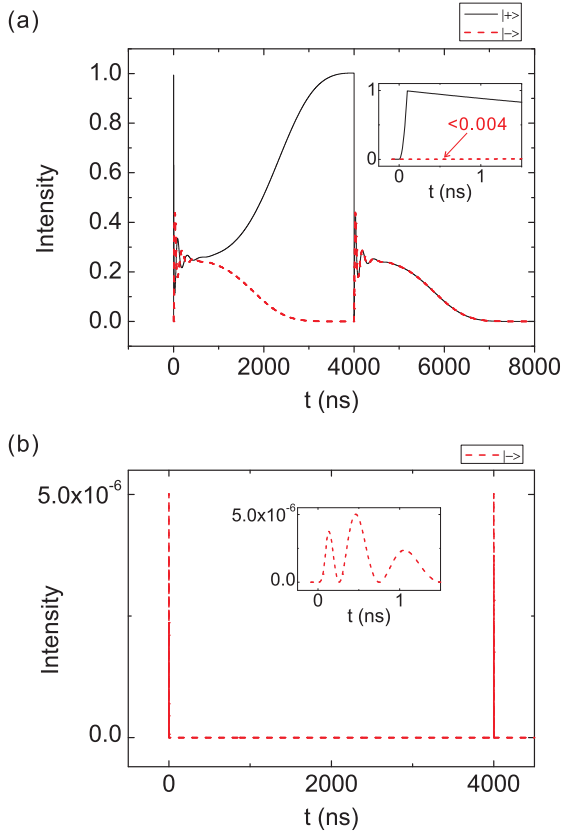}
\end{center}
\caption{ (color online). Intensity of the transmitted light of the polarizations $|+\rangle$ (black solid line) and $|-\rangle$ (red dashed line)
without the filter (a) and with the filter (b).}
\end{figure}

To show the advantages of having the filter present, the simulation results using FFT with 0.01-ns
sampling resolution are shown in Fig. 3 with $z_{1}=0.01$ m and $z_{2}=0$ for panel (a) and $z_{1}=0.01$ m and
$z_{2}=0.01$ m for panel (b). The input pulse is the same as before and $\delta_{2}$ was chosen to be
$150\delta_{1}$. For Fig. 3(a), we first note that the $V$ polarized light remains the same as the input
square pulse after traveling in the pure YVO$_4$ crystal as mentioned above. When the $H$ polarized
wavefronts decay, the $V$ polarized light starts to dominate. Therefore, $I_{+}$ (black solid line)
and $I_{-}$ (red dashed line) have nearly the same intensity. Then, the $H$ polarized main fields
slowly increase. $I_{+}$ rises with it and $I_{-}$ decays, because the delayed main fields are
still almost entirely in the phase with the $V$ polarized light. After 4 $\mu$s, the $V$ polarized pulse ends,
and $I_{+}$ and $I_{-}$ share exactly the same intensities. It can be seen that $I_{+}$ rises
immediately after $t=0$, but $I_{-}$ is suppressed during the first nanosecond. This shows that $H$
polarized wavefronts travel at the same speed as the $V$ polarized light in the pure YVO$_{4}$
crystal. If this were not the case, $I_{-}$ would rise at the same time and with the same intensity
as $I_{+}$, because only $H$ ($V$) polarized light would present and there would be no interference between
the two paths for the given delay time. A time delay $\tau_{d}$ between the arrival of $H$ and $V$
polarized light would lead to a remarkable intensity for $I_{-}$, which can be estimated to be
$\tau_{d}/t_{r}$. However, for Fig. 3(a) with no filter present, $I_{-}$ can grow to 0.004 in the
first nanoseconds with no delay for either path. Therefore, it becomes impossible to measure
$\tau_{d}$ with subpicosecond accuracy.

The transmission $I_{-}(z_{2},t)$ with the filter is shown in Fig. 3(b). $I_{+}$ is the same as
that in Fig. 3(a) and is not shown in Fig. 3(b). Because the filter has wideband absorption, only
photons with frequencies far-detuned from the resonance are transmitted. The main fields are
totally absorbed by the filter, and only rising and falling edges at $t=0$ and $t=t_{0}$ remain.
For all time scales, at most $I_{-}$ can grow to $5\times 10^{-6}$ in the first nanoseconds after
$t=0$ and $t=t_{0}$. Because the remaining far-detuned light experiences little phase shift, there
is near-perfect interference between the two paths. Therefore, whenever there is an intensity
larger than $10^{-5}$ recorded for $I_{-}$, we may infer that there is a time delay between the
arrival of $H$ and $V$ polarized light. Moreover, the interference yields the same results when the
first Nd$^{3+}$:YVO$_{4}$ crystal acts as a two-level system, which is the fast-light regime.

\begin{figure}[tbph]
\begin{center}
\includegraphics[width= 3.1in]{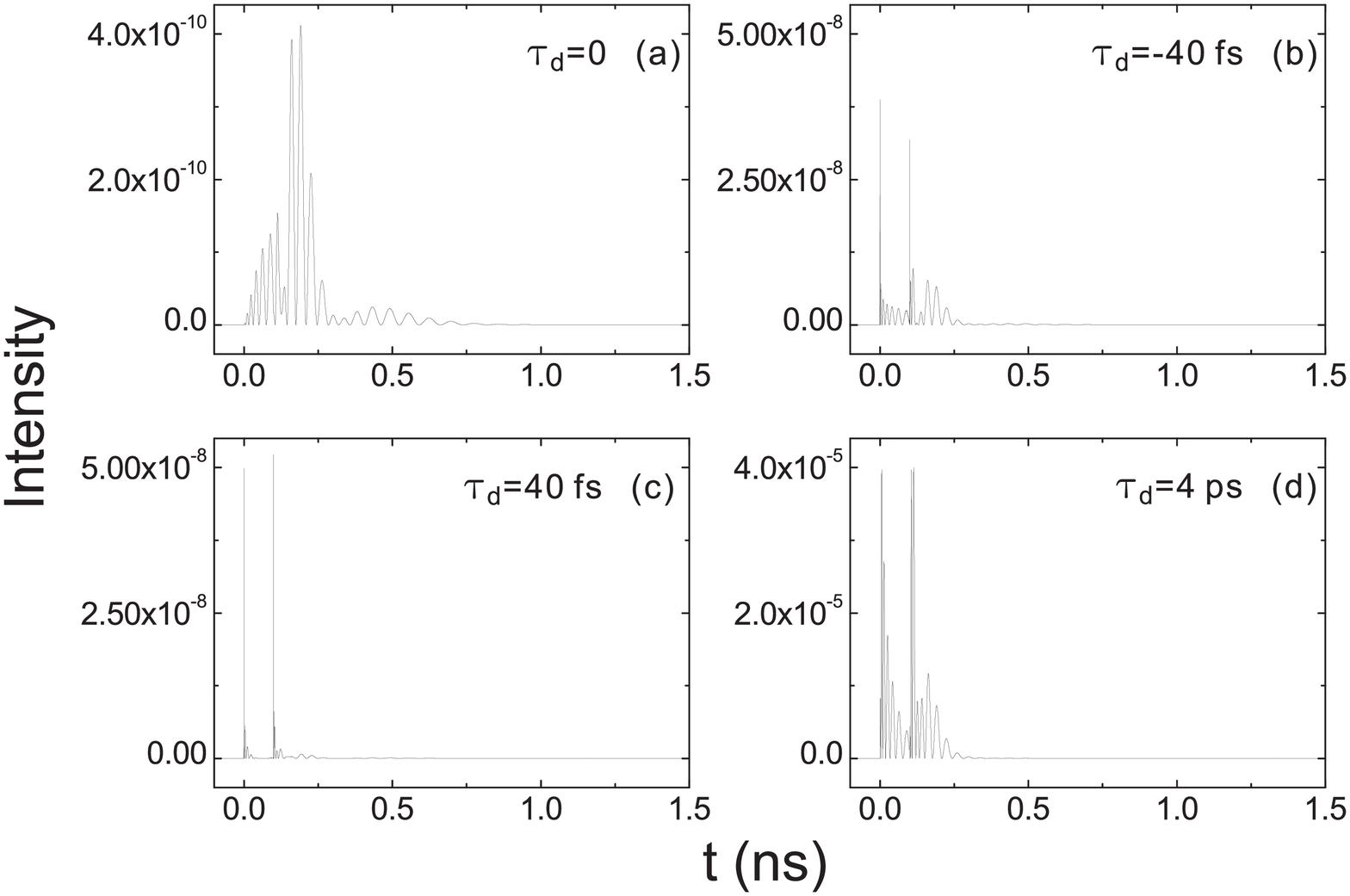}
\end{center}
\caption{ Intensity of $|-\rangle$ polarized light for $\tau_{d}=0$ (a), $\tau_{d}=$ -40 fs (b), $\tau_{d}=$ 40 fs (c) and $\tau_{d}=$ 4 ps (d).}
\end{figure}

To accurately determine the speed of wavefronts using this method, let $z_{1}=0.055$ m,
$z_{2}=0.08$ m, and $\delta_{2}=1050\delta_{1}$ which is roughly the inhomogeneous broadening of
the Nd$^{3+}$ ions. Thus, the lower AOM in Fig. 2 may not be necessary. The propagation time in the
first crystal is $t_{1}\simeq z_{1}/\upsilon_{0}\simeq$ 400 ps. A time delay $\tau_{d}$ between the
$H$ polarized light and $V$ polarized light is artificially introduced. Let
$E_{H}(z_{2},\omega)=\sqrt{1/2}E(0,\omega)Exp[ik_{1}(\omega)z_{1}]Exp[i\omega\tau_{d}]$. This means
the $H$ polarized light will fall behind (move ahead of) of the $V$ polarized light by $|\tau_{d}|$ for
$\tau_{d}>0$ ($\tau_{d}<0$). The output intensities calculated using FFT with 0.1-ps sampling
resolution with $\tau_{d}=$ 0, -40 fs, 40 fs and 4 ps are shown in Figs. 4(a)-(d), respectively. Clearly
$I_{-}$ with $|\tau_{d}|=$ 40 fs is much greater than that without delay. The longer the delay, the
greater the $I_{-}$ that can be obtained. Whether the $H$ polarized light is faster or slower than
the $V$ polarized light has little effect on the output. This is shown in Fig.s 4(b) and 4(c).
The sign of $\tau_{d}$ may be experimentally identified by introducing a compensation delay in
either path.

Considering a realistic detector with a bandwidth of 5 GHz, the recorded power for $|-\rangle$
polarized light can be estimated by averaging the results in Fig. 4(b) for every 0.2 ns. The
maximum outputs are $9.2\times10^{-11}, 2.5\times10^{-9}, 5.9\times10^{-10}$ and
 $6.6\times10^{-6}$ for Figs. 4(a)-(d), respectively. The power of probe light can be chosen to be 1 mW in the experiment.
Therefore, when the power of $|-\rangle$ polarized light is recorded with a value over 2.5 pW, we
can conclude that there is a speed difference of $\Delta$$\upsilon> \upsilon_{0}/10^{4}$ for the two
paths, which is approximately $\Delta$$\upsilon/\upsilon_{0}\simeq \tau_{d}/t_{1}$. With the
parameters chosen here, a bound of less than 10$^{-4}$ on the accuracy of the wavefronts' speed can
be experimentally determined.

The highest speed resolution with this scheme is not limited to $10^{-4}$ which depends on several
parameters. The first is the rise time of the input pulse, which is determined by the modulation
method. Because $I_{-}\propto \tau_{d}/t_{r}$, the smaller $t_{r}$ is, the more sensitive $I_{-}$
is to $\tau_{d}$. The faster the signal rises, the harder they can ``see" Nd$^{3+}$ in the
Nd$^{3+}$:YVO$_{4}$ crystal. The second is the length of the first crystal, $z_{1}$, which
determines the transmission time, $t_{1}$. Because $\Delta$$\upsilon/\upsilon_{0}\simeq
\tau_{d}/t_{1}$, for a larger $t_{1}$, the speed difference can be determined more precisely.
Third, detectors with better low-intensity responses can resolve a smaller $\tau_{d}$. Note that
the rise time of the detectors is not so important in this scheme because only the maximal
intensity is of interest rather than the absolute time that the signal rises. Therefore, this method
can determine the speed of the precursor to an accuracy of a given lower bound with better
experimental conditions. Compared with previous works which determine the speed of information to an
accuracy of about 0.1 by direct intensity recording \cite{Gauthier05,Gauthier03}, our scheme
shall give more accurate measurements.

To summarize, we propose an interference scheme for accurately measuring the speed of wavefronts in
an optically pumped Nd$^{3+}$:YVO$_{4}$ crystal. The simulation results computed with linear
susceptibility of EIT medium show that the arrival time of wavefronts is independent of the strong
dispersions caused by Nd$^{3+}$ ions. Considering the current available modulation and detection
technologies, the pure YVO$_4$ crystal is treated as an infinite fast-response medium like the
vacuum, so the wavefronts' speed we are measuring is $\upsilon_{0}$.  In principle, with a powerful
wideband filter and single-photon detectors with extremely low noise, the precursors which travel
at the speed of $c$ in a Nd:YVO$_4$ crystal also can be identified with our scheme. This scheme can be
carried out in other systems where no background wideband medium is present, such as cold
atoms and room temperature atoms. The conclusion can be made general as the ideal step-rising
wave fronts travel at exactly the same speed $c$ in all materials. This scheme provides an approach
to accurately measure the arrival time of precursors in various systems with realistic detecting
systems. Future experimental measurements will provide a strict test of whether the information
velocity with medium violates the Einstein causality. This work bridges the gap between the
ideal concepts of precursors and the real pulses with finite rise (fall) time by showing an accurate
boundary.

This work was supported by the National Basic Research Program (2011CB921200), National Natural
Science Foundation of China (Grant Nos. 60921091 and 10874162).

\end{document}